\documentclass[aps,prd,twocolumn,showpacs,amssymb,footinbib]
{revtex4}

\usepackage{amsmath,amsthm,amsfonts}

\begin{document}

\title{Loop quantization of vacuum Bianchi I cosmology}
\author{M. Mart\'{i}n-Benito}
\email{merce.martin@iem.cfmac.csic.es}
\affiliation{Instituto de Estructura de la Materia,
CSIC, Serrano 121, 28006 Madrid, Spain}
\author{G. A. Mena Marug\'{a}n}\email{mena@iem.cfmac.csic.es}
\affiliation{Instituto de Estructura de la Materia,
CSIC, Serrano 121, 28006 Madrid, Spain}
\author{T. Pawlowski}\email{tomasz@iem.cfmac.csic.es}
\affiliation{Instituto de Estructura de la Materia,
CSIC, Serrano 121, 28006 Madrid, Spain}

\begin{abstract}

We analyze the loop quantization of the family of
vacuum Bianchi I spacetimes, a gravitational system
whose classical solutions describe homogeneous
anisotropic cosmologies. We rigorously construct the
operator that represents the Hamiltonian constraint,
showing that the states of zero volume completely
decouple from the rest of quantum states. This fact
ensures that the classical cosmological singularity is
resolved in the quantum theory. In addition, this
allows us to adopt an equivalent quantum description
in terms of a well defined densitized Hamiltonian
constraint. This latter constraint can be regarded in
a certain sense as a difference evolution equation in
an internal time provided by one of the triad
components, which is polymerically quantized.
Generically, this evolution equation is a relation
between the projection of the quantum states in three
different sections of constant internal time.
Nevertheless, around the initial singularity the
equation involves only the two closest sections with
the same orientation of the triad. This has a double
effect: on the one hand, physical states are
determined just by the data on one section, on the
other hand, the evolution defined in this way never
crosses the singularity, without the need of any
special boundary condition. Finally, we determine the
inner product and the physical Hilbert space employing
group averaging techniques, and we specify a complete
algebra of Dirac observables. This completes the
quantization program.

\end{abstract}


\maketitle

\section{Introduction}

Loop quantum cosmology (LQC) \cite{lqc} is nowadays an
active field of research, devoted to the application
of the ideas and mathematical methods of the full
theory of Loop Quantum Gravity (LQG)
\cite{lqg3,lqg1,lqg2} to symmetry reduced cosmological
models. This application is useful at least in two
respects. On the one hand, it allows us to learn and
gain experience about issues that are still open in
full LQG. On the other hand, in many cases the
symmetry reduced models already give us information
about physical questions of interest. The first
attempts to apply the techniques of LQG to symmetry
reduced models can be found in Ref. \cite{boj1}. More
recently, some homogeneous and isotropic models have
been quantized to completion in the LQC framework
\cite{aps1,aps2,aps3,kale,apsv,skl,vand,tom} along the
revisited lines presented in Ref. \cite{abl}. In
particular, these studies provide new results about
the fate of the classical singularities. Namely, the
cosmological singularities are resolved dynamically in
these models, as they are replaced with quantum
bounces.

In this paper we will discuss the loop quantization of
a homogeneous but anisotropic model: the Bianchi I
spacetimes in vacuo. Some preliminary analyses on the
quantization of the Bianchi I model using Ashtekar
variables were already developed in Refs.
\cite{aspu,neg}. The merit for the first systematic
attempts to construct the kinematical Hilbert space
and introduce a Hamiltonian constraint for the model
in a loop quantization framework must be granted to
Bojowald \cite{boj}. However, apart from technical
issues concerning the definition of the quantum
operators (and the prescription adopted to incorporate
the presence of a gap in the area spectrum of LQG),
the analysis of Ref. \cite{boj} was not complete
inasmuch as it did not provide the physical Hilbert
space, nor an algebra of Dirac observables. The first
work that attempted to complete the Dirac quantization
program, adapting the techniques presented in Ref.
\cite{abl} to quantize polymerically the gravitational
degrees of freedom of the Bianchi I spacetimes, was
done by Chiou \cite{chio}. In that case, nonetheless,
the considered homogeneous model was not in a vacuum,
because it included a massless scalar field. Here we
will employ the same kind of techniques although we
will use a slightly different quantization
prescription \cite{thesis}, which seems more suitable
to make manifest some relevant aspects of the LQC
approach, keep under rigorous control the definition
of the quantum Hamiltonian constraint, and complete
the analysis of the physical states. In particular, an
important feature of the quantization proposed here is
that it immediately leads to the decoupling of the
quantum states with zero volume \cite{F0}, so that
they can be removed from the theory. Employing this
fact, we will show that physical states can be
described equivalently as solutions to a densitized
Hamiltonian constraint. We will prove that the
operator that represents this constraint is a well
defined self-adjoint operator. Furthermore, we will
explicitly construct the solutions to the constraint,
find the physical inner product, and determine a
complete set of observables.

Most of the homogenous models analyzed so far in LQC
\cite{aps1,aps2,aps3,kale,apsv,skl,vand,tom}, and in
particular the mentioned work of Ref. \cite{chio},
contain matter in the form of a homogeneous massless
scalar field. This field, quantized in a standard
(Schr\"odinger-like) way serves as an internal time,
providing a well defined notion of evolution. Here, in
the vacuum case, such a useful object is no longer
available. Hence we will explore the problem of
quantizing the variable that plays the role of
internal time in a polymeric way, a fact that may
affect the conventional concept of evolution. In our
case, we will choose as internal time one of the triad
components. This choice will provide us with a certain
notion of evolution, as we will discuss in Sec. VII.

The classical vacuum Bianchi I spacetimes generically
possess an initial cosmological singularity, to which
we will allocate the origin of time. One of the main
motivations of our analysis is to discuss what happens
with this singularity in the loop quantum theory. We
will see that, since the zero-volume states are
totally decoupled, non-trivial physical states contain
no contribution from them. In this sense the initial
classical singularity is resolved and disappears from
the quantum theory. Furthermore, in the presented
quantization there exists no correlation between the
sectors of opposite orientations of the triads.
Therefore, the defined evolution does not connect
sectors corresponding to different orientations of the
variable identified as the internal time. As a
consequence, the singularity is not crossed and no
additional branch of the universe emerges on the
opposite side of it. The provided notion of evolution
is well defined without the need to impose any special
boundary condition to deal with the two orientations
of the triads.

The polymeric quantization of the system leads to a
densitized Hamiltonian constraint that can be viewed
as a discrete evolution equation on quantum states.
This equation is a recurrence relation, which
generically relates the projection of the state in
three consecutive sections of constant internal time.
However, when one reaches the origin in the
quantization presented in this paper, it reduces to a
relation on the two sections with the smallest
possible values of the discrete time, in the sector of
triad orientations under study. As a result, the
physical states are in fact determined by their data
on a single section. The vector space of these data
can be provided with an inner product and one attains
in this way the physical Hilbert space.

An additional, important motivation for the analysis
of the loop quantization of Bianchi I in vacuo comes
from the consideration of a (much richer) family of
cosmological spacetimes that contain inhomogeneities,
namely, the linearly polarized Gowdy model with
$T^3$-topology \cite{let}. This is an infinite
dimensional model that provides a most suitable arena
for the attempt to extend the analysis of homogenous
LQC to inhomogeneous situations. The subfamily of
homogeneous solutions within this model is just the
classical vacuum Bianchi I spacetimes with
$T^3$-topology. From this perspective, the loop
quantization of Bianchi I in vacuo is a preliminary
step in order to face the quantization of the Gowdy
cosmologies.

Let us also comment that, during the writing of this
manuscript, another work on the loop quantization of
Bianchi I has appeared \cite{luc} which presents some
similarities with our treatment. That work, carried
out independently to ours, considers a simplified
version of the quantization, where the corrections
owing to the regularization of the inverse triad
operator are not incorporated, and describes the time
evolution using a massless scalar field, like in Ref.
\cite{chio}.

The main body of this paper is organized as follows.
In Sec. II we construct the kinematical Hilbert space
on which we define the elementary operators of the
theory. The Hamiltonian constraint of the model is
represented as a symmetric operator in Sec. III, where
we also show the decoupling of the zero-volume states.
Employing this decoupling, we densitize the
Hamiltonian constraint in Sec. IV, arriving to an
equivalent quantum description of the system. The form
of the densitized constraint allows us to decompose it
in terms of one-dimensional operators, which are then
analyzed in detail in Sec. V. The solutions to the
constraint and the corresponding physical Hilbert
space is determined in Sec. VI. Finally, in Sec. VII
we discuss the results of our quantization and
conclude.

\section{Kinematics}

As a first step towards the loop quantization of the
vacuum Bianchi I spacetimes, we describe the model in
terms of Ashtekar variables \cite{chio,chi2}. In
principle, the definition of these variables makes use
of a finite sized cell and a fiducial triad. Adopting
a diagonal gauge, it was shown in Ref. \cite{chi2}
that there is no physical dependence on the choice of
fiducial triad if one defines the homogenous canonical
variables for the model in a suitable way. For the
sake of simplicity, we will then particularize the
discussion to the choice of a diagonal Euclidean
triad. The issue of the dependence on the coordinate
cell of integration is more subtle \cite{chi2}.
Nonetheless, when the spatial sections of the Bianchi
I cosmologies have a compact topology, there is a
natural choice of coordinate cell. In fact, as we have
mentioned in the Introduction, one of the motivations
for our study is the potential application to the
quantization of the homogeneous sector of the linearly
polarized $T^3$-Gowdy model \cite{let}. Consequently,
we will specialize our analysis to a compact
three-torus topology, adopting the corresponding
natural cell for our treatment, namely the $T^3$-cell
with sides of coordinate length equal to $2\pi$. In
this way, one arrives to the following non-trivial
components of the $SU(2)$ gravitational connection
$A_i^a$ and of the densitized triad $E_a^i$ \cite{F1},
\begin{equation}
A_i^a=\frac{c^{i}}{2\pi}\delta_i^a, \qquad
E_a^i=\frac{p_{i}}{4\pi^2}\delta_a^i,
\end{equation}
so that $\{c^i,p_j\}=8\pi G\gamma\delta^i_j$. Here
$i,j=1,2,\text{ or }3$ are spatial indices, $a$ is an
internal $SU(2)$ index, $G$ is the Newton constant,
and $\gamma$ is the Barbero-Immirzi parameter.

The spacetime metric written in terms of the variables
$p_i$ takes the form: \begin{equation} ds^2= -N^2
dt^2+\frac{|p_1p_2p_3|}{4\pi^2}\left[ \sum_{i=1}^3
\frac{(dx^i)^2}{p_i^2}\right],\end{equation} where
$\{dx^i\}$ is the fiducial co-triad, with $x^i\in
S^1$, and $N$ is the lapse function.

In LQC, one adapts the techniques of LQG to symmetry
reduced systems in order to construct the kinematical
Hilbert space \cite{abl,boj,chio}. The configuration
variables are provided by holonomies along edges
oriented in the fiducial directions, and the momentum
variables by triad fluxes through fiducial rectangles
orthogonal to those directions. The holonomy along an
edge of oriented coordinate length $2\pi\mu_i$ in the
direction $i$ is
\begin{equation}
h_i^{\mu_i}(c^i)=e^{ \mu_{i}c^{i}\tau_{i}},
\end{equation}
where $\tau_i$ are the $SU(2)$ generators proportional
to the Pauli matrices, such that
$[\tau_i,\tau_j]=\epsilon_{ijk}\tau^k$. The flux
through the rectangle of coordinate area $S^{i}$
orthogonal to the direction $i$ turns out to be
\begin{equation}
E[S^{i}]=\frac{p_{i}}{4\pi^2}S^{i}.
\end{equation}

The configuration algebra is obtained from the sums of
products of matrix elements of the irreducible
representations of the holonomies, and is just the
algebra of almost periodic functions of $c^i$
\cite{abl,chio}. This algebra is generated by the
exponentials
\begin{equation}
\mathcal N_{\mu_i}(c^i)=e^{\frac{i}{2}\mu_{i}c^{i}},
\end{equation}
which, using the Dirac ket notation, will be
represented by the states $|\mu_i\rangle$. The finite
linear combinations of products of these functions
provide the analog of the space of cylindrical
functions in LQG, and we will call it $\text{Cyl}_S$.
Thus, denoting
$|\mu_1,\mu_2,\mu_3\rangle=\otimes_i|\mu_i\rangle$, we
have
\begin{equation}
\text{Cyl}_S=\text{span}\{|\mu_1,\mu_2,\mu_3\rangle\}.
\end{equation}

The kinematical Hilbert space $\mathcal
H_{\text{Kin}}=\otimes_i \mathcal H_{\text{Kin}}^i$ is
the completion of the space $\text{Cyl}_S$ with
respect to the discrete inner product
$\langle\mu_i|\mu_i^\prime\rangle=\delta_{\mu_i
\mu_i^\prime}$ for each direction \cite{abl,chio}. The
states $|\mu_i\rangle$ provide an orthonormal basis
for $\mathcal H_{\text{Kin}}^i$. They are eigenstates
of the operator $\hat p_i$ associated to fluxes, while
$\hat{\mathcal N}_{\mu_i^\prime}$ simply shifts their
label $\mu_i$:
\begin{eqnarray}
\hat p_i|\mu_i\rangle&=&4\pi\gamma
l_\text{Pl}^2\mu_i|\mu_i\rangle,
\\
\hat{\mathcal
N}_{\mu_i^\prime}|\mu_i\rangle&=&|\mu_i+\mu_i^\prime
\rangle.\end{eqnarray} Here, $l_\text{Pl}=\sqrt{G
\hbar}$ is the Planck length.

In LQG, the operator that represents the physical area
has a discrete spectrum, with a minimum nonzero
eigenvalue equal to $\Delta=2\sqrt{3}\pi\gamma
l_{\text{Pl}}^2$. It has been argued that, when one
takes into account this fact, a minimum coordinate
length for the edge of the holonomies is introduced in
LQC \cite{aps3}. The exact form in which such a
minimum coordinate length must be incorporated is
still under discussion. At present, two prescriptions
are considered in the literature \cite{chi2,ref}. Here
we will adopt the prescription introduced in Ref.
\cite{chio}, usually called the $\bar{\mu}$-scheme.
One of the advantages of this prescription is that (as
we will see) the quantum analysis of the system can be
carried out to completion, and not just in an
effective, non-fundamental way. Besides, this will
allow us to revisit some parts of the analysis
presented in Ref. \cite{chio} which, up to date, is
the most complete discussion of the loop quantization
of the Bianchi I model. In doing so, we will see that
one can also learn some lessons about the quantization
of other homogeneous systems like the isotropic ones.
On the other hand, although the justification of this
prescription from the viewpoint of the full theory of
LQG is currently under investigation \cite{PVA}, it is
important to note that there are no inconsistencies or
non-physical effects associated to it in cases with
compact spatial topology like the one considered here,
cases when a privileged coordinate cell exists
\cite{F2}.

According to this $\bar\mu$-scheme, the minimum
coordinate length for each direction $i$ is determined
by the condition $\bar{\mu}_i^2 |p_i|=\Delta$, from
which we arrive at the operator relation
\begin{equation}\label{muop}
\widehat{\frac{1}{\bar\mu_i}}=
\frac{\widehat{\sqrt{|p_i|}}}{\sqrt{\Delta}}.
\end{equation}
Operators like $\widehat{\sqrt{|p_i|}}$ are defined in
terms of $\hat{p}_i$ by means of the associated
spectral decomposition. Acting on a state
$|\mu_i\rangle$, we then get
\begin{equation}
\widehat{\frac{1}{\bar\mu_i}}|\mu_i\rangle=
\frac{1}{\bar\mu_i(\mu_i)}|\mu_i\rangle,
\quad\bar\mu_i(\mu_i)=
\sqrt{\frac{\sqrt{3}}{2|\mu_i|}}.
\end{equation}
Since the value $\bar\mu_i$ is state-dependent, the
associated operator $\hat{\mathcal N}_{\bar\mu_i}$
generates a state-dependent minimum shift. To write
down its action, it is convenient to relabel the
states by reparametrizing $\mu_i$ so that the minimum
shift becomes uniform. This is achieved by
introducing, for each direction, a label $v_i(\mu_i)$
which satisfies the equation
\begin{equation}
\bar\mu_i(\mu_i)\frac{\partial}{\partial\mu_i}
=\frac{\partial}{\partial v_i},
\end{equation}
whose solution is
\begin{equation}
v_i(\mu_i)=\sqrt{\frac{2^3}{3^{5/2}}}
\text{sgn}(\mu_i)|\mu_i|^{3/2}.
\end{equation}
With this relabeling, the basic operators have the
following action in the domain $\text{Cyl}_S$:
\begin{eqnarray}
\hat p_i|v_i\rangle&=&3^{1/3}\Delta\,\text{sgn}(v_i)
|v_i|^{2/3}|v_i\rangle,\\ \label{nop}\hat{\mathcal
N}_{\bar\mu_i}|v_i\rangle&=&|v_i+1\rangle.
\end{eqnarray}

\section{Hamiltonian constraint}

In the considered model, only one constraint remains
to be imposed once the diagonal gauge has been chosen:
the Hamiltonian constraint. In order to represent it
as an operator, one first needs to express this
constraint as a phase space function in terms of
triads and holonomies, since there do not exist well
defined operators corresponding to the connection
components $c^i$. This is done by the standard
procedures of LQC, explained in detail in Ref.
\cite{chio}. In brief, one defines the curvature
components employing holonomies along edges of
coordinate length $2\pi \bar{\mu}_i$ and regularizes
the inverse of the determinant of the metric following
Thiemann's procedure \cite{lqg3}, i.e., expressing it
via the Poisson bracket of holonomies with the volume
function. Applying these procedures and setting the
lapse $N=1$, we arrive to the following form for the
Hamiltonian constraint \cite{chio} (integrated over
the chosen $T^3$-cell and still viewed as a classical
function on phase space)
\begin{eqnarray}\label{CBI}
C_{BI}&=&-\frac{2}{\gamma^2} \left[\Lambda_1\Lambda_2
\left(\frac{1}{\sqrt{|p_3|}}\right)_{\text{reg}}+
\Lambda_1\Lambda_3
\left(\frac{1}{\sqrt{|p_2|}}\right)_{\text{reg}}\right.
\nonumber\\
& +& \left. \Lambda_2\Lambda_3
\left(\frac{1}{\sqrt{|p_1|}}\right)_{\text{reg}}
 \right],
\end{eqnarray}
where
\begin{equation}
\Lambda_i= \frac{\sqrt{|p_i|}}{\bar{\mu}_i}
\text{sgn}(p_i)\sin(\bar\mu_ic^i),
\end{equation}
and $(1/\sqrt{|p_i|})_{\text{reg}}$ is the regularized
expression for $1/\sqrt{|p_i|}$ obtained via
Thiemann's method.

By the mentioned standard LQC procedures, this
regularized phase space function is represented by the
operator
\begin{align}
\widehat{\left[\frac{1}{\sqrt{|p_i|}}\right]}&=
\frac1{4\pi\gamma
l_{\text{Pl}}^2}\widehat{\frac{1}{\bar\mu_i}}
\widehat{\text{sgn}(p_i)}\nonumber\\
&\times\left(\hat{\mathcal
N}_{-\bar\mu_i}\widehat{\sqrt{|p_i|}}\hat{\mathcal
N}_{\bar\mu_i}-\hat{\mathcal
N}_{\bar\mu_i}\widehat{\sqrt{|p_i|}}\hat{\mathcal
N}_{-\bar\mu_i}\right),\label{preg}
\end{align}
where, for the $\bar{\mu}$-scheme that we have
adopted, $\widehat{1/\bar\mu_i}$ and $\hat{\mathcal
N}_{\bar\mu_i}$ are the operators defined in Eqs.
(\ref{muop}) and (\ref{nop}), respectively. Note that
there is no factor ordering ambiguity in the above
formula, inasmuch as the operator in parenthesis
commutes with all the others on the right-hand side of
Eq. (\ref{preg}). Besides, it is easy to check that
the states $|v_i\rangle$ are eigenstates of the
introduced operator. Explicitly,
\begin{align}
&\widehat{\left[\frac{1}{\sqrt{|p_i|}}\right]}
|v_i\rangle=b(v_i)|v_i\rangle,\nonumber\\
&b(v_i)=\frac{3^{5/6}}{2\sqrt{\Delta}}
|v_i|^{1/3}\left||v_i+1|^{1/3}-|v_i-1|^{1/3}\right|.
\end{align}

In order to construct a symmetric operator
$\widehat{C}_\text{BI}$ representing the Hamiltonian
constraint \eqref{CBI}, let us now consider the
quantum counterpart of $\Lambda_i$. From Eq.
(\ref{muop}), it follows that all the factors in
$\Lambda_i$ depend only on $p_i$ except for
$\sin(\bar\mu_ic^i)$. This latter term can be
represented by the operator
\begin{equation}
\widehat{\sin(\bar\mu_ic^i)}=\frac{1}{2i}(\hat{\mathcal
N}_{2\bar\mu_i}-\hat{\mathcal N}_{-2\bar\mu_i}),
\end{equation}
which does not commute with $\hat{p}_i$. To obtain a
symmetric operator for $\Lambda_i$, we then proceed as
follows. Since the operator
\begin{equation}
\widehat{\sqrt{|p_i|}}\widehat{\frac{1}{\bar{\mu}_i}}=
\frac{1}{\sqrt{\Delta}}\widehat{|p_i|}\end{equation}
is non-negative, we can take its square root and adopt
the symmetric factor ordering
\begin{align}\label{Lambdasym}
\widehat{\Lambda}_i=
\frac{1}{2\sqrt{\Delta}}\widehat{\sqrt{|p_i|}}
&\bigg[\widehat{\sin(\bar\mu_ic^i)}
\widehat{\text{sgn}(p_i)}\nonumber\\
&+\widehat{\text{sgn}(p_i)}
\widehat{\sin(\bar\mu_ic^i)}\bigg]
\widehat{\sqrt{|p_i|}}.
\end{align}

Several comments are in order at this point. First, it
is important to realize the presence of the factor
$\text{sgn}(p_i)$ in the expression of $\Lambda_i$,
which was not properly taken into account in Ref.
\cite{chio}. As a function on phase space, it does not
commute with $\sin(\bar\mu_ic^i)$ under Poisson
brackets, and hence their product as operators is not
symmetric. Probably, its appearance had not been
pointed out so far because, in the passage to the best
studied case of (homogeneous and) isotropic LQC, a
simplification occurs that makes its role less
important. Up to a constant factor, the purely
gravitational part of the Hamiltonian constraint for
isotropic models can be obtained from Eq. (\ref{CBI})
by identifying the three spatial directions
\cite{aps3,chio}. In doing so, this gravitational part
gets a factor of a squared sign that can be considered
equal to the unity and ignored, instead of dealing
with it as we have discussed for the anisotropic case.
These two alternatives for the isotropic models can be
understood as different choices of factor ordering. We
have checked that a factor ordering like the one
suggested here does not alter significantly the
numerical results of Refs. \cite{aps2,aps3} (in fact,
for situations of physical interest, the difference is
below the numerical errors). Nonetheless, even for the
isotropic case our factor ordering may be more
convenient in order to clarify certain conceptual and
technical issues, like e.g. the decoupling of the
zero-volume states or the properties of the solutions
to the constraint near the cosmological singularity,
as we will see later on.

In addition, it is important to notice that the
operator $\widehat{\Lambda}_i$ annihilates the state
$|v_i=0\rangle$, which belongs to the kernel of
$\hat{p}_i$. Furthermore, the range of
$\widehat{\Lambda}_i$ does not contain the alluded
state, so that its orthogonal complement in ${\cal
H}^i_{\text{Kin}}$ is invariant under the action of
$\widehat{\Lambda}_i$. In particular, this ensures
that the action of the operator
$\widehat{\text{sgn}(p_i)}$ present in
$\widehat{\Lambda}_i$ is well defined \cite{F3}.

The factor ordering adopted in the quantum Hamiltonian
constraint has some important advantages with respect
to that proposed in Ref. \cite{chio}. First, the
constraint is now a sum of products of symmetric
operators, each defined on one of the Hilbert spaces
${\cal H}^i_{\text{Kin}}$ associated to each
direction. As we will see, this facilitates the
determination of observables and makes the
construction of physical solutions straightforward,
thus allowing to complete the quantization. Second, it
is easy to check that the Hamiltonian constraint
annihilates the proper subspace $\mathcal
H_{\text{Kin}}^0$ of states in the kernel of any of
the operators $\hat{p}_i$. Such subspace is the
completion of the subset of ${\text{Cyl}}_S$ given by
${\text{Cyl}}_S^0=\text{span}\{|v_1,v_2,v_3\rangle;\,
v_1v_2v_3=0\}$. Since $\hat
V=\otimes_i\widehat{\sqrt{|p_i|}}$ is the volume
operator, we will call $\mathcal H_\text{Kin}^0$ the
subspace of zero-volume states. Furthermore, the
properties of the operator $\widehat{\Lambda}_i$
commented above imply that the orthogonal complement
of $\mathcal H_{\text{Kin}}^0$ is invariant under the
action of the constraint $\widehat{C}_{\text{BI}}$.
Thus, the subspace of zero-volume states decouples
from its complement and we can ignore it in the
following, restricting our considerations exclusively
to the subspace of nonzero-volume states. We will call
this subspace $\widetilde{\mathcal H}_\text{Kin}$,
whereas $\widetilde{\text{Cyl}}_S$ will denote the
corresponding linear span of tensor products of states
$|v_i\rangle$ such that none of the $v_i$'s vanishes.

As we will discuss in the next section, the decoupling
of the zero-volume states allows one to describe the
quantum system in a completely equivalent way in terms
of a densitized version of the Hamiltonian constraint.
Moreover, since non-trivial physical states get no
contribution from zero-volume states, the classical
initial singularity disappears from the quantum
theory, already at the kinematical level. At least in
this sense, the singularity is resolved quantum
mechanically, in a way similar to that originally
suggested by Bojowald \cite{boj} (see nonetheless
\cite{F4}). We will consider this issue in more detail
in the last section.

\section{Densitized Hamiltonian constraint}

In order to solve the quantum constraint, it proves
convenient to recast it in a densitized form which is
easier to analyze. One should remember that physical
states are states annihilated by the Hamiltonian
constraint $\widehat{C}_{\text{BI}}$ and, in
principle, they do not have to be normalizable in the
kinematical Hilbert space $\widetilde{\mathcal
H}_\text{Kin}$. More precisely, we expect these states
to live in a larger space, namely the algebraic dual
$\widetilde{\text{Cyl}}_S^*$ of the dense set
$\widetilde{\text{Cyl}}_S$. We will denote one such
state by $(\psi|$.

In order to densitize the quantum Hamiltonian
constraint in a rigorous manner, we have to invert the
action of the operator
\begin{equation}\label{vinv}
\widehat{\left[\frac{1}{V}\right]}= \otimes_i
\widehat{\left[\frac{1}{\sqrt{|p_i|}}\right]}
\end{equation}
which (via Thiemann's procedure) entered in the
definition of our constraint. At this point, the
observation that the zero-volume states decouple is
essential, because the kernel of the operator
(\ref{vinv}) coincides precisely with that subspace.
Thus, the inverse operator ${\widehat{[1/V]}\,^{-1}}$
is well defined once we have restricted ourselves to
$\widetilde{\mathcal H}_\text{Kin}$. Note also that
this densitization can be carried out exactly, without
the need to simplify the theory by ignoring the
quantum corrections coming from the regularized
inverse volume operator (i.e., without replacing
${\widehat{[1/V]} \,^{-1}}$ by $\hat{V}$).

To reformulate the constraint in its densitized
version, we introduce the following bijection in the
dual $\widetilde{\text{Cyl}}_S^*$
\begin{equation}
(\psi|\longrightarrow(\psi|
\widehat{\left[\frac{1}{V}\right]}^{\frac1{2}}.
\end{equation}
The transformed physical states are now annihilated by
the (adjoint of the) symmetric densitized Hamiltonian
constraint $\widehat{{\cal C}}_{\text{BI}}$, defined
as
\begin{equation}
\widehat{{\cal C}}_{\text{BI}}=
\widehat{\left[\frac1{V}\right]}^{-\frac1{2}}
\widehat{C}_{\text{BI}}\widehat{\left[\frac{1}{V}
\right]}^{-\frac1{2}}.
\end{equation}
Its explicit form is
\begin{equation}\label{C}
\widehat{{\cal
C}}_{\text{BI}}=-\frac{2}{\gamma^2}\bigg[\widehat{\Theta}_1
\widehat{\Theta}_2+\widehat{\Theta}_1\widehat{\Theta}_3+
\widehat{\Theta}_2\widehat{\Theta}_3\bigg],
\end{equation}
where $\widehat{\Theta}_i$ is the symmetric operator
\begin{equation}
\widehat{\Theta}_i=
\widehat{\left[\frac{1}{\sqrt{|p_i|}}\right]}^{-\frac1{2}}
\widehat{\Lambda}_i
\widehat{\left[\frac1{\sqrt{|p_i|}}\right]}^{-\frac1{2}}.
\end{equation}
This operator has the following action on the basis
states $|v_i\rangle$:
\begin{equation}\label{acttheta}
\widehat{\Theta}_i|v_i\rangle=-i\frac{\Delta}{2\sqrt{3}}
\big[f_+(v_i)|v_i+2\rangle-f_-(v_i)|v_i-2\rangle\big],
\end{equation}
where
\begin{eqnarray}\label{f}
f_\pm(v_i)&=&g(v_i\pm2)s_\pm(v_i)g(v_i), \\\label{s}
s_\pm(v_i)&=&\text{sgn}(v_i\pm2)+\text{sgn}(v_i),
\end{eqnarray}
and
\begin{align}\label{g}
g(v_i)&=&
\begin{cases}
\left|\left|1+\frac1{v_i}\right|^{\frac1{3}}
-\left|1-\frac1{v_i}\right| ^{\frac1{3}}
\right|^{-\frac1{2}} & {\text{if}} \quad v_i\neq 0,\\
0 & {\text{if}} \quad v_i=0.\\
\end{cases}
\end{align}

\section{Analysis of the constraint operator}

One of the advantages of our quantization procedure is
that, in order to study the properties of the
constraint operator $\widehat{{\cal C}}_{\text{BI}}$,
we only need to analyze the operator
$\widehat{\Theta}_i$ on $\widetilde{\cal
H}_{\text{Kin}}^i$. We carry out that analysis in this
section.

\subsection{Superselection}

As we see in Eq. \eqref{acttheta},
$\widehat{\Theta}_i$ is a difference operator with a
step of two units in the label $v_i$. Given the
definitions \eqref{f}-\eqref{g}, the function
$f_+(v_i)$ vanishes in the whole interval
$v_i\in[-2,0]$, while $f_-(v_i)$ is equal to zero for
$v_i\in[0,2]$. Owing to this remarkable property,
which can be traced back to our treatment of the
factor $\text{sgn}(p_i)$ in the constraint, the
operator $\widehat\Theta_i$ does not relate states
$|v_i\rangle$ with $v_i>0$ to those with $v_i<0$.
Therefore, $\widehat\Theta_i$ connects only states
with labels $v_i$ belonging to one of the semilattices
\begin{equation}
\mathcal
L_{\varepsilon_i}^\pm=\{\pm(\varepsilon_i+2k),k\in
\mathbb{N}\},\end{equation} where
\begin{equation}
\mathbb{N}=\mathbb{N}^+\cup\{0\},
\quad\varepsilon_i\in(0,2].
\end{equation}

Semilattices corresponding to different values of
$\varepsilon_i$ or to different signs are not
connected by the action of $\widehat\Theta_i$. In
other words, the Hilbert space $\mathcal
H_{\varepsilon_i}^{\pm}$, defined as the Cauchy
completion of the set
\begin{equation}
\text{Cyl}_{\varepsilon_i}^{\pm}=\text{span}\{|v_i\rangle;
v_i\in\mathcal L_{\varepsilon_i}^\pm\}
\end{equation}
with respect to the discrete inner product, is
invariant under the action of $\widehat\Theta_i$. Note
that the kinematical Hilbert space for each direction,
which is not separable, can be decomposed into these
separable Hilbert spaces:
\begin{equation}
\widetilde{\mathcal
H}_{\text{Kin}}^i=\oplus_{\varepsilon_i}(\mathcal
H_{\varepsilon_i}^+\oplus\mathcal
H_{\varepsilon_i}^-).
\end{equation}

Because of the absence of physically relevant
operators that connect the different semilattices, the
physical Hilbert space is then divided into
superselection sectors. We can thus restrict our study
to any specific Hilbert space $\mathcal
H_{\vec\varepsilon}^+=\otimes_i\mathcal
H_{\varepsilon_i}^+$, with
$\vec\varepsilon=(\varepsilon_1,\varepsilon_2,
\varepsilon_3)$. Equivalently, we could construct the
theory using e.g. the Hilbert space ${\mathcal
H}_{\vec\varepsilon}^-=\otimes_i\mathcal
H_{\varepsilon_i}^-$, since the constraint is
symmetric under a flip of sign in the label $v_i$,
owing to the identity
\begin{equation}\label{symsi}
f_\pm(-v_i)=-f_\mp(v_i).
\end{equation}

Finally, we would like to emphasize that the state
$|v_i=0\rangle$ is not included in any of the
superselection sectors, since it had been removed from
the kinematical Hilbert space. The semilattices which
one might expect to be connected with this state under
the action of the constraint are those corresponding
to $\varepsilon_i=2$, but one can explicitly check
that $\widehat{\Theta}_i$ is indeed a completely well
defined operator on $\mathcal H_{\varepsilon_i}^{\pm}$
with domain $\text{Cyl}_{\varepsilon_i}^{\pm}$.
Nothing special occurs in the case $\varepsilon_i=2$
in comparison with the other possible superselection
sectors.

\subsection{Spectral analysis}

In order to determine the spectral properties of the
operator $\widehat\Theta_i$ it is helpful to first
analyze its square. This squared operator is also
important by itself as it represents (up to a
multiplicative constant) the gravitational part of the
densitized constraint in the isotropic case, where the
elementary variables corresponding to the three
different spatial directions are identified.

One can easily check that $\widehat\Theta_i^2$ is a
difference operator of constant step equal to four in
the label $v_i$. Its action couples only those points
$v_i$ which lay on one of the semilattices
\begin{equation}
^{(4)}{\mathcal L}_{\tilde\varepsilon_i}^\pm=
\{\pm(\tilde\varepsilon_i+4k), k\in\mathbb{N}\},\quad
\tilde\varepsilon_i\in(0,4].
\end{equation}
Thus, $\widehat\Theta_i^2$ leaves invariant each of
the Hilbert spaces $^{(4)}{\mathcal
H}_{\tilde\varepsilon_i}^{\pm}$ obtained by the
completion of
\begin{equation}
^{(4)}\text{Cyl}_{\tilde\varepsilon_i}^{\pm}=
\text{span}\{|v_i\rangle;\,v_i\in {^{(4)}{\mathcal
L}_{\tilde\varepsilon_i}^\pm}\}.
\end{equation}

If one now defines $\widehat\Theta_i^2$ in
$^{(4)}{\mathcal H}_{\tilde\varepsilon_i}^{+}\oplus
{^{(4)}{\mathcal H}_{4-\tilde\varepsilon_i}^{-}}$
(with domain
$^{(4)}\text{Cyl}_{\tilde\varepsilon_i}^{+}\cup
{^{(4)}\text{Cyl}}_{4-\tilde\varepsilon_i}^{-}$), it
is not difficult to check that its difference with
respect to the operator $H_\text{APS}^\prime
\Delta^2/(\pi G)$ defined in Ref. \cite{kale} [see Eq.
(37) in that reference] is just a symmetric, trace
class operator. For the particular case
$\tilde\varepsilon_i=4$, we can establish the same
kind of comparison with $H_\text{APS}^\prime$ by
starting with the Hilbert space $^{(4)}{\mathcal
H}_{4}^{+}\oplus {^{(4)}{\mathcal H}_{4}^{-}}$ and
then including the state $|v_i=0\rangle$, defining
e.g. a vanishing action of $\widehat\Theta_i^2$ on it.

Using the results obtained in Ref. \cite{kale} about
the operator $H_\text{APS}^\prime$ and Kato's
perturbation theory \cite{kato}, it is straightforward
to prove that $\widehat\Theta_i^2$ is a positive,
essentially self-adjoint operator whose essential
spectrum and absolutely continuous spectrum are
$[0,\infty)$ \cite{RS,GP}.

On the other hand, since $\widehat\Theta_i^2$ leaves
invariant $^{(4)}{\mathcal
H}_{\tilde\varepsilon_i}^{\pm}$, its restriction to
$^{(4)}{\mathcal
H}_{\tilde\varepsilon_i}^{+}\oplus{^{(4)}{\mathcal
H}_{4-\tilde\varepsilon_i}^{-}}$ (which has been
analyzed above) commutes e.g. with the projection onto
the subspace $^{(4)}{\mathcal
H}_{\tilde\varepsilon_i}^{+}$. As a consequence, we
conclude that $\widehat\Theta_i^2$ on the Hilbert
space $^{(4)}{\mathcal H}_{\tilde\varepsilon_i}^{+}$
(with domain
$^{(4)}\text{Cyl}_{\tilde\varepsilon_i}^{+}$) is
essentially self-adjoint. Otherwise its deficiency
index equation would have non-trivial solutions that
would provide also valid solutions for the case in
which the operator is defined on the larger Hilbert
space $^{(4)}{\mathcal
H}_{\tilde\varepsilon_i}^{+}\oplus {^{(4)}{\mathcal
H}_{4-\tilde\varepsilon_i}^{-}}$, reaching a
contradiction because we have already established that
the operator is essentially self-adjoint in this
latter case. Besides, for $^{(4)}{\mathcal
H}_{\tilde\varepsilon_i}^{+}$, the essential spectrum
and the absolutely continuous spectrum must still be
$[0,\infty)$. One can show it taking into account the
symmetry of $\widehat\Theta_i^2$ under a flip of sign
in the label $v_i$ [see Eq. \eqref{symsi}] and
accepting the independence of the spectrum on the
value of $\tilde{\varepsilon}_i$. Moreover, numerical
studies \cite{mmp} indicate that the whole spectrum is
just absolutely continuous. Indeed, the spectrum of
$\widehat\Theta_i^2$ (with domain
$^{(4)}\text{Cyl}_{\tilde\varepsilon_i}^{+}$) is
non-degenerate and each of its eigenfunctions
converges for large $v_i$ to a non-vanishing
eigenfunction of the geometrodynamical
(Wheeler-DeWitt) counterpart of the operator. This,
together with the continuity of the spectrum in
geometrodynamics, suffices to conclude that the
discrete and singular spectra are empty \cite{WK}.

In a manner similar to that explained above, one can
also relate the solutions to the deficiency index
equation of the operators $\widehat\Theta_i^2$ and
$\widehat\Theta_i$ on ${\mathcal
H}_{\varepsilon_i}^{+}={^{(4)}{\mathcal
H}_{\varepsilon_i}^{+}}\oplus {^{(4)}{\mathcal
H}_{2+\varepsilon_{i}}^{+}}$ (both with domain
$\text{Cyl}_{\varepsilon_i}^{+}$). In this way, one
can deduce that the specified operator
$\widehat\Theta_i$ is essentially self-adjoint.
Therefore, we conclude that the constraint operator
$\widehat{{\cal C}}_{\text{BI}}$, given in Eq.
\eqref{C} and defined in the domain
$\text{Cyl}_{\vec\varepsilon}^+= \otimes_i
\text{Cyl}_{\varepsilon_i}^+$, is in fact essentially
self-adjoint.

\subsection{Generalized eigenstates}

Taking into account the results of the previous subsection, we can
obtain the spectral resolution of the identity associated to the
operator $\widehat\Theta_i$, e.g. on ${\mathcal
H}_{\varepsilon_i}^{+}$, starting with those for the squared
operator $\widehat\Theta_i^2$ on ${^{(4)}{\mathcal
H}_{\varepsilon_i}^{+}}$ and ${^{(4)}{\mathcal
H}_{2+\varepsilon_i}^{+}}$. Remember that the spectrum of
$\widehat\Theta_i^2$ on any of these two Hilbert spaces is
absolutely continuous and equal to the positive real line. For all
$\rho_i$ in the spectrum, we will call
$|{^{(4)}e_{\rho_i}^{\tilde\varepsilon_i}})$ the corresponding
generalized eigenstates normalized to the Dirac delta \cite{F5},
where $\tilde\varepsilon_i=\varepsilon_i$ or $2+\varepsilon_i$.
Thus, on ${^{(4)}{\mathcal H}_{\tilde\varepsilon_i}^{+}}$ we have
\begin{equation}
\mathbb{I}=\int_{\mathbb{R^+}} d\rho_i
|{^{(4)}e_{\rho_i}^{\tilde\varepsilon_i}})
({^{(4)}e_{\rho_i}^{\tilde\varepsilon_i}}|.
\end{equation}
In addition, we fix the global phase of these
generalized eigenstates by choosing
$({^{(4)}e_{\rho_i}^{\tilde\varepsilon_i}}|v_i\rangle$
to be a positive number for $v_i
=\tilde\varepsilon_i$. In particular, this choice and
the positivity of the operator $\widehat\Theta_i^2$
ensure that
$({^{(4)}e_{\rho_i}^{\tilde\varepsilon_i}}|v_i\rangle$
is real for all $v_i\in{}^{(4)}{\mathcal
L}_{\tilde\varepsilon_i}^{+}$.

Renaming $\rho_i=\omega_i^2$ and combining the above
resolutions of the identity for the Hilbert spaces
${^{(4)}{\mathcal H}_{\varepsilon_i}^{+}}$ and
${^{(4)}{\mathcal H}_{2+\varepsilon_i}^{+}}$, we
obtain that, on their direct sum ${\mathcal
H}_{\varepsilon_i}^{+}$,
\begin{equation}
\mathbb{I}=\int_{\mathbb{R}} d\omega_i
|e_{\omega_i}^{\varepsilon_i})
(e_{\omega_i}^{\varepsilon_i}|,
\end{equation}
where, for $\omega_i\neq 0$,
\begin{eqnarray}
|e_{+|\omega_i|}^{\varepsilon_i})&
=&\sqrt{|\omega_i|}\left[
|{^{(4)}e_{\omega_i^2}^{\varepsilon_i}})-
i|{^{(4)}e_{\omega_i^2}^{2+\varepsilon_i}})
\right],\nonumber\\
|e_{-|\omega_i|}^{\varepsilon_i})&=&\sqrt{|\omega_i|}
\left[|{^{(4)}e_{\omega_i^2}^{\varepsilon_i}})+
i|{^{(4)}e_{\omega_i^2}^{2+\varepsilon_i}})\right].
\end{eqnarray}
For $\omega_i=0$, we define \begin{equation}
|e_{0}^{\varepsilon_i})=|{^{(4)}e_{0}^{\varepsilon_i}}).
\end{equation}
Recalling Eq. \eqref{acttheta}, one can check that the
states $|e_{\omega_i}^{\varepsilon_i})$ defined above
are generalized eigenstates of $\widehat\Theta_i$ on
${\mathcal H}_{\varepsilon_i}^{+}$, with $\omega_i$
being the corresponding eigenvalue. Whence we find a
real, absolutely continuous spectrum. Note also that
the projections of $|e_{\omega_i}^{\varepsilon_i})$ on
the Hilbert subspaces ${^{(4)}{\mathcal
H}_{\varepsilon_i}^{+}}$ and ${^{(4)}{\mathcal
H}_{2+\varepsilon_i}^{+}}$ (which are generalized
eigenstates of the squared operator
$\widehat\Theta_i^2$ except for $\omega_i=0$, when one
of the projection vanishes) have a relative phase of
$\pm\pi/2$. As a consequence, the phase of
\begin{equation}\label{ev}
e_{\omega_i}^{\varepsilon_i}(v_i)= \langle
v_i|e^{\varepsilon_i}_{\omega_i})\end{equation}
oscillates rapidly when $v_i$ varies in the
semilattice ${\mathcal L}_{\varepsilon_i}^+$.

Furthermore, using Eq. \eqref{acttheta}, the
eigenvalue equation associated to the operator
$\widehat\Theta_i$ on ${\mathcal
H}_{\varepsilon_i}^{+}$ leads to the recurrence
equation
\begin{align}\label{eigrec}
e^{\varepsilon_i}_{\omega_i}&(2n+2+\varepsilon_i)
=\frac{g(2n-2+\varepsilon_i)}{g(2n+2+\varepsilon_i)}
e^{\varepsilon_i}_{\omega_i}(2n-2+\varepsilon_i)
\nonumber\\&-i\frac{\sqrt{3}\omega_i}{\Delta}
\frac{e^{\varepsilon_i}_{\omega_i}(2n+\varepsilon_i)}
{g(2n+2+\varepsilon_i)g(2n+\varepsilon_i)}
\end{align}
$\forall n\in\mathbb{N}^+$, which involves three
distinct values of $v_i$, as it corresponds to a
second-order difference equation. However, for $n=0$
we get a relation between the two first coefficients
of the generalized eigenstates,
\begin{equation}\label{consist}
e^{\varepsilon_i}_{\omega_i}(2+\varepsilon_i)=
-i\frac{\sqrt{3}\omega_i}{\Delta}
\frac{e^{\varepsilon_i}_{\omega_i}
(\varepsilon_i)}{g(2+\varepsilon_i)
g(\varepsilon_i)}.
\end{equation}
Therefore the solutions to the eigenvalue problem are
totally determined by a simple piece of initial data,
namely, the projection of the generalized eigenstate
for the minimum allowed value of $v_i$,
$e^i_{\omega_i}(\varepsilon_i)$. Actually, a careful
calculation shows that, $\forall n\in \mathbb{N}^+$,
\begin{align}
e^{\varepsilon_i}_{\omega_i}&(2n+\varepsilon_i)=
\sum_{O(n)} \left[\prod_{\{r_p\}}
\frac{g(2r_p+\varepsilon_i)}{g(2r_p+4+\varepsilon_i)}\right]
\nonumber\\
&\times \left[\prod_{\{s_q\}}
\frac{-i\sqrt{3}\omega_i}{\Delta
g(2s_q+2+\varepsilon_i)g(2s_q+\varepsilon_i)}\right]
e^{\varepsilon_i}_{\omega_i}(\varepsilon_i).
\label{solemio}\end{align} Here, $O(n)$ denotes the
set of all possible ways to move from 0 to $n$ by
jumps of one or two steps. For each element in $O(n)$,
$\{r_p\}$ is the subset of integers followed by a jump
of two steps, whereas $\{s_q\}$ is the subset of
integers followed by a jump of only one step.

\section{The physical Hilbert space}

Once we have a good knowledge of the properties of the
constraint operator, we turn to the issue of
determining the physical Hilbert space in order to
complete the quantization. Let ${\cal U}$ be the
domain of the self-adjoint extension of the constraint
operator $\widehat{{\cal C}}_{\text{BI}}$ on the
superselection sector ${\mathcal
H}_{\vec\varepsilon}^+$. Starting from the dense
subset ${\cal U}$, we can obtain the physical Hilbert
space, $\mathcal H^\text{Phy}_{\vec\varepsilon}$, by
applying the group averaging procedure
\cite{gave,gave2}. Namely, given an element
$|\phi\rangle$ in ${\cal U}$ [with corresponding wave
function $\phi(\vec{v})$ in the
$\vec{v}$-representation, where
$\vec{v}=(v_1,v_2,v_3)$], one can ``project'' it onto
$\mathcal H^\text{Phy}_{\vec\varepsilon}$ via an
average over the group generated by the constraint
$\widehat{{\cal C}}_{\text{BI}}$:
\begin{equation}\label{eq:gave}
\Phi(\vec{v}) = [{\cal P} \phi](\vec{v})=
\int_{\mathbb{R}} d t\,
e^{it\frac{\gamma^2}{2}\widehat{{\cal C}}_{\text{BI}}}
\phi(\vec{v}).
\end{equation}

Employing the spectral decomposition associated to the
operators $\widehat{\Theta}_i$ for the three spatial
directions, we can express the wave function
$\phi(\vec{v})$ in terms of the eigenfunctions
$e_{\omega_i}^{\varepsilon_i}(v_i)$ introduced in Eq.
\eqref{ev}
\begin{equation}\label{eq:fdec}
\phi(\vec{v}) = \int_{\mathbb{R}^3} d\vec{\omega} \,
\bar{\phi}(\vec{\omega})
e^{\varepsilon_1}_{\omega_1}(v_1)
e^{\varepsilon_2}_{\omega_2} (v_2)
e^{\varepsilon_3}_{\omega_3}(v_3) ,
\end{equation}
where $\vec{\omega}=(\omega_1,\omega_2,\omega_3)$,
$\bar{\phi}\in L^2(\mathbb{R}^3,d\vec{\omega})$, and
$v_i\in\mathcal L_{\varepsilon_i}^+$. Substituting
this decomposition into Eq. \eqref{eq:gave}, we then
get
\begin{align}\label{sol}
\Phi(\vec{v}) &= \int_{\mathbb{R}^3}d\vec{\omega}
\delta(\omega_1\omega_2+\omega_1\omega_3+
\omega_2\omega_3)\bar{\phi}(\vec{\omega})\nonumber\\
&\times e^{\varepsilon_1}_{\omega_1}(v_1)
e^{\varepsilon_2}_{\omega_2}(v_2)
e^{\varepsilon_3}_{\omega_3}(v_3).
\end{align}

We immediately conclude from this expression that only
products of eigenstates
$|e^{\varepsilon_i}_{\omega_i})$ with
$\sum_{i}(1/\omega_i)=0$ will contribute to $\mathcal
H^\text{Phy}_{\vec\varepsilon}$. It is then useful to
consider one of the $\omega_i$'s as a function of the
others. Owing to the symmetry of the system under the
interchange of the three directions, the particular
choice of $\omega_i$ that one makes is just a matter
of convention. Here we select $\omega_1$. From now on,
we define it as follows:
\begin{equation}\label{om-func}
\omega_1(\omega_2,\omega_3)=
-\frac{\omega_2\omega_3}{\omega_2+\omega_3}.
\end{equation}
With this choice, the wave function that represents
the ``projection'' of the kinematical state
$|\phi\rangle$ onto the physical Hilbert space takes
the form
\begin{align}
\Phi(\vec{v}) &= \int_{\mathbb{R}^2} \frac{d\omega_2
d\omega_3}{|\omega_2+\omega_3|}\,
\Phi(\omega_2,\omega_3)\,
e^{\varepsilon_1}_{\omega_1(\omega_2,\omega_3)}(v_1)
\nonumber\\
&\times e^{\varepsilon_2}_{\omega_2} (v_2)
e^{\varepsilon_3}_{\omega_3}(v_3) ,\label{Phisol}
\end{align}
where $\Phi(\omega_2,\omega_3)=
\bar{\phi}[\omega_1(\omega_2,\omega_3),\omega_2,\omega_3]$.
We remember that the eigenfunctions
$e^{\varepsilon_i}_{\omega_i}(v_i)$ that appear in the
above expression are given (up to a normalization
factor) by Eq. \eqref{solemio}.

Employing this result and the delta-normalization of
the generalized eigenstates of $\widehat{\Theta}_i$,
one can compute the physical inner product between two
states $|\Phi_1\rangle$ and $|\Phi_2\rangle$
\begin{align}
\langle \Phi_1|\Phi_2\rangle_\text{Phy} &= \langle
{\cal P}\phi_1|\phi_2\rangle_\text{Kin}
\nonumber\\
&= \int_{\mathbb{R}^2} \frac{d\omega_2
d\omega_3}{|\omega_2+\omega_3|}\,
\Phi^{*}_1(\omega_2,\omega_3)
\Phi_2(\omega_2,\omega_3).
\end{align}
We have introduced the subindices ``$\text{Phy}$'' and
``$\text{Kin}$'' to distinguish between the inner
products of the physical and the kinematical Hilbert
spaces, and the symbol $^{*}$ denotes complex
conjugation. Therefore, the physical Hilbert space of
the considered system is the space
\begin{equation}\label{physpa}\mathcal
H^\text{Phy}_{\vec\varepsilon}= L^2\left(\mathbb{R}^2,
\frac{d\omega_2
d\omega_3}{|\omega_2+\omega_3|}\right)\end{equation}
of square integrable functions on $\mathbb{R}^2$ with
integration measure $d\omega_2
d\omega_3/|\omega_2+\omega_3|$.

An alternative way to arrive to the physical Hilbert
space is to find the space of solutions to the
constraint and a complete set of (real) observables,
imposing then the condition that these observables be
self-adjoint in order to determine a Hilbert structure
on the space of solutions. Once we know the spectral
resolution of the identity associated to the operators
$\widehat{\Theta}_i$, it is straightforward to solve
the constraint by adopting a formal expansion of the
states in terms of the generalized eigenstates
$|e_{\omega_i}^{\varepsilon_i})$, since
$\widehat{{\cal C}}_\text{BI}$ has a diagonal action
with this decomposition. Indeed, if we represent the
candidate state by a wave function
$\phi'(\omega_1,\omega_2,\omega_3)$, $\widehat{{\cal
C}}_\text{BI}$ becomes just a polynomial constraint on
$\omega_i$. Hence, the physical solutions are
described by functions of the form
$\Phi'(\omega_2,\omega_3)=\phi'[\omega_1(\omega_2,\omega_3),
\omega_2,\omega_3]$, where
$\omega_1(\omega_2,\omega_3)$ is given by Eq.
\eqref{om-func}. A complete set of observables is
provided e.g. by the operators $\widehat{\Theta}_2$
and $\widehat{\Theta}_3$, that multiply the wave
function $\Phi'(\omega_2,\omega_3)$ by $\omega_2$ and
$\omega_3$ respectively, and by the derivative
operators $-i\partial_{\omega_2}$ and
$-i\partial_{\omega_3}$. Demanding that they be
self-adjoint, we arrive to the physical Hilbert space
$L^2(\mathbb{R}^2,d\omega_2 d\omega_3)$. Under
multiplication of the wave function by the factor
$\sqrt{|\omega_2+\omega_3|}$, one obtains in fact a
unitarily equivalent representation of the algebra of
observables on the Hilbert space given in Eq.
\eqref{physpa}.

\section{Conclusions and discussion}

In this work, we have presented a complete loop
quantization of the family of homogeneous Bianchi I
cosmologies in vacuo. We have described the quantum
system in terms of a well defined densitized
Hamiltonian constraint, represented it by a(n
essentially) self-adjoint operator, found the general
form of its solutions, and determined the
corresponding Hilbert space of physical states.

A nice property of the quantization that we have put
forward, including the chosen factor ordering, is that
the quantum Hamiltonian constraint adopts the form of
a sum of products of symmetric operators for each of
the spatial directions. Under the densitization of the
constraint, this allows one to reduce the analysis of
self-adjointness and the spectral analysis just to the
consideration of operators for one-dimensional systems
(those describing independently each of the spatial
directions).

Furthermore, our quantum Hamiltonian constraint
annihilates the kernel of the volume operator and
leaves invariant its orthogonal complement. As a
consequence, the subspace of zero-volume states gets
decoupled and one can ignore it in the study of the
non-trivial physical states. This fact is essential to
attain an equivalent formulation of the quantum system
in terms of a densitized Hamiltonian constraint. This
densitized constraint retains all the information
about the quantum behavior of the model; in
particular, it is not necessary to introduce
simplifications that disregard the regularization of
the inverse triad operators in LQC.

As it is common in polymeric quantizations, the
physical Hilbert space is superselected. Physical
states have support in semilattices that have a basic
step of two units in the labels $v_i$. This differs
from other previous analysis in LQC where the
constructed superselection sectors correspond to
entire lattices, with points distributed over the
whole real line instead of over a semiaxis
\cite{abl,aps1,aps2,aps3,kale,apsv,skl,vand,tom,boj,chio,luc}.
As we have seen in Sec. V, this facilitates the
spectral analysis of the relevant difference operators
and removes remnant global symmetries of the system,
namely those under reversal of the triad orientations
(which now simply relate different superselection
sectors). In addition, and related with these issues,
all superselection sectors have essentially the same
treatment in our quantization. Nothing special happens
for the semilattices formed by even integers, which
are those that the difference operators would connect
with the origin had the zero-volume states not got
decoupled.

As we have commented, the factor ordering that we have
adopted reduces the functional analysis of the
densitized constraint operator to that of the simple
operator $\widehat{\Theta}_i$. In addition, for the
three possible directions $i$, these operators are
immediately identified as Dirac observables. It is
worth commenting some of their properties in more
detail. We have seen that $\widehat\Theta_i$ behaves
as a second-order difference operator in many aspects.
Nonetheless, we know that it represents the classical
variable $c^ip_i$, which would become a first-order
differential operator in a standard canonical
quantization of the Wheeler-DeWitt type. Therefore,
one would expect that, if the results of the
Wheeler-DeWitt approach are to be recovered in a
certain regime from the loop quantization, the
generalized eigenstates have to be determined from
data on a single section of constant $v_i$.
Remarkably, this is indeed the case.

This issue had not been pinpointed until now because
in the homogeneous and isotropic models a subtle
coincidence occurs. In that case, the gravitational
part of the constraint is given by an operator like
$\widehat\Theta_i^2$. Such an operator, which
represents the classical variable $(c^i p_i)^2$,
behaves in fact as a second-order one in the
Wheeler-DeWitt quantization.

Another difference with respect to previous analyses
in LQC like those presented in Refs.
\cite{abl,aps1,aps2,aps3,kale,apsv,skl,vand,tom,chio,luc}
is that we have not introduced any matter content (and
specifically no scalar field) as internal time. The
variable identified with the internal time in our
model is one of the triad components, which are
quantized polymerically. Actually, the three diagonal
triad components have an equivalent role because the
model is symmetric under the interchange of spatial
directions. We have selected the direction $i=1$ to
make the discussion explicit, but the choice is just a
question of convention. Since the eigenstates of
$\widehat\Theta_1$ are determined by their initial
data at $v_1=\varepsilon_1$, the physical states get
also fixed by data on this initial slice. Furthermore,
the non-degeneracy of the spectrum of
$\widehat\Theta_1$ implies that the wave function at
any $v_1$-constant slice determines the entire
solution up to corrections of vanishing physical norm.
Also the physical inner product can be rewritten in
terms of data on that single section
\cite{ip-nonl-foot}. One can then admit the viewpoint
that physical states evolve from this initial slice to
any other slice, in such a way that they solve the
densitized Hamiltonian constraint. As a consequence of
the loop quantization adopted for the internal time,
this concept of evolution differs from the usual one
and, in particular, does not allow to reach unitarity
in a straightforward, standard way \cite{mmp}. One
should be aware that the notion of evolution used here
refers only to the fact that there is a deterministic
relation between data on two slices of constant
internal time, and thus its meaning is certainly
limited. Also, a preliminary analysis of the
eigenfuctions shows that the evolution parameter
(internal time) is not monotonic in the cosmic time,
which makes the extraction of relevant physical data
non-trivial.

On the other hand, the decoupling of the zero-volume
states ensures that non-trivial physical states have
no contribution from the slice where the curvature
singularity is located in the classical theory.
Physical states are well defined everywhere and, in
this respect, the singularity is resolved quantum
mechanically. This conclusion reinforces previous
results about singularity resolution in LQC. It is
worth emphasizing that this conclusion is independent
of our choice of internal time. Furthermore, since the
wave functions of the physical states (in the $v_1$
representation) have support just in semilattices, the
evolution does not connect them with other branches of
the universe, corresponding to a different orientation
of the triads. The singularity is never crossed in the
evolution. Let us also notice that this result is
achieved without appealing to any kind of boundary
conditions that might restrict the initial data for
the physical states. As far as one understands the
statement in this sense, one can say that physical
states ``arise from nothing'' in the initial slice
$v_1=\varepsilon_1$, attaining a no-boundary
description.

Apart from the points addressed in this work about the
loop quantization of the vacuum Bianchi I model, there
are other interesting issues that we plan to explore
in a future research \cite{mmp}. This includes a
detailed comparison with the Wheeler-DeWitt theory,
the discussion of the concept of evolution, and the
fate of unitarity in the model. We also want to carry
out a numerical study of the physical states,
analyzing in particular the occurrence of quantum
bounces.

\section*{Acknowledgments}

The authors are thankful to A. Ashtekar, M. Bojowald,
A. Corichi, L.J. Garay, and J.M. Velhinho for
enlightening conversations and discussions. This work
was supported by the Spanish MEC Project
FIS2005-05736-C03-02 (with its continuation
FIS2008-06078-C03-03) and the Spanish
Consolider-Ingenio 2010 Programme CPAN
(CSD2007-00042). M.M.-B. and T.P. acknowledge
financial support by the I3P Program of CSIC and the
European Social Fund, in the case of M.M.-B. under the
grant I3P-BPD2006.


\begin{thebibliography}{99}
\bibitem{F1} We do not employ the Einstein convention
for summation of indices.
\bibitem{F2} For non-compact topology, it was seen in Ref.
\cite{chi2} that the effective dynamics arising from
this prescription presents a dependence on the
coordinate cell, assuming that such an effective
dynamics emerges exactly as in the homogeneous and
isotropic models \cite{effec,effec2}. Even in this
situation, one may find a slight modification of the
prescription so that the commented dependence
disappears, replacing Eq. (B14) in Ref. \cite{chi2}
with $\bar{\mu}_i^2 |p_i| L_j L_k=\Delta$.
\bibitem{F3} The action of
$\widehat{\text{sgn}(p_i)}$ on the state
$|v_i=0\rangle$ can be defined as the multiplication
by an arbitrary constant inasmuch as the final action
of $\widehat{\Lambda}_i$ is independent of that
choice.
\bibitem{F4} However, as we will see, the singularity is
never crossed in the quantum evolution, contrary to
the suggestions of Ref. \cite{boj}. In addition,
preliminary numerical analyses indicate that the
singularity is also resolved dynamically in the
physically interesting cases.
\bibitem{F5} We employ a slightly modified bra-ket notation
to make clear that these states do not belong to the considered
kinematical Hilbert space.
\bibitem{ip-nonl-foot} At this point it is worth
noticing that, given the data on one $v_1$-constant
slice, the physical inner product is non-local when
expressed in terms of $v_2$ and $v_3$. Partly related
to this it is the fact that the space of physical
states is smaller than the space of solutions that one
would expect for the three-dimensional difference
equation associated with the Hamiltonian constraint.
This can be seen by considering the formulation of the
initial value problem: in principle, to find the
unique solution one would need to specify the data on
two of the three initial surfaces $v_i=\varepsilon_i$
(for example $i=1,2$). On the other hand, to determine
the physical state we just need the data on one
initial slice.
\end{thebibliography}
\end{document}